\documentclass[twocolumn,showpacs,preprintnumbers,amsmath,amssymb]{revtex4}
\usepackage{graphicx}% Include figure files
\usepackage{dcolumn}% Align table columns on decimal point
\usepackage{bm}% bold mat

\begin{document}

%\preprint{APS/123-QED}

\title{Size scaling of  the addition spectra in  silicon quantum dots}% Force line breaks with \\

\author{M.~Boehm, M.~Hofheinz, X.~Jehl, M.~Sanquer}
\email{msanquer@cea.fr}
\affiliation{CEA-DRFMC, 17 rue des Martyrs, F-38054 Grenoble, France}
\author{M.~Vinet, B.~Previtali, D.~Fraboulet, D.~Mariolle, 
S.~Deleonibus}
\affiliation{CEA-DRT-LETI, 17 rue des Martyrs F-38054 Grenoble Cedex 9, France}

\date{\today}% It is always \today, today,

\begin{abstract}
We investigate small artificial quantum dots obtained by geometrically controlled resistive confinement in low mobility silicon-on-insulator nanowires. Addition spectra were recorded at low temperature for various dot areas fixed by lithography. We compare the standard deviation of the addition spectra with theory in the high electron concentration regime. We find that the standard deviation scales as the inverse area of the dot and its absolute value is comparable to the energy spacing of the one particle spectrum.
\end{abstract}

\pacs{73.23.-b, 73.23.Hk, 73.63.-b}% PACS, the Physics and Astronomy
\maketitle
%
%\section{\label{sec:level1}Introduction:\protect\\}
%

Measuring the current as a function of gate voltage provides decisive information on the addition spectrum of electrons in quantum dots. The addition energy is given by 
$\Delta_{2}(N)=E_{N+1}+E_{N-1}-2E_{N}= e\alpha \delta V_{g}(N)$, where $\delta V_{g}(N)$ is the difference in gate voltage between the $N^{th}$ and the $(N+1)^{th}$ current resonance, $\alpha = {C_{g}\over C_{\sum}}$ the ratio between the gate and total capacitance of the dot and $E_{N}$ the ground-state energy of the $N$-electron state \cite{alhassid}.
Few experiments concentrate on the fluctuations of the addition energy, $\sigma_2  = \sqrt{\langle\Delta_{2}^{2}\rangle -\langle\Delta_{2}\rangle^{2}}$, which require a large number of resonances to be measured \cite{patel,simmel,simmel2,sivan,luscher}.   
The distribution of $\Delta_{2}$ found in experiments is generally close to a Gaussian with a standard deviation $\sigma_2$ varying from values comparable to the mean one-particle level spacing $\Delta_{1}$ to much larger values. Until now the statistical variation of addition spectra were studied in dots made with high mobility  2D electron gases with a large  $k_{F} \ell $ parameter, where $k_{F} $ is the Fermi wavevector and $ \ell $ the elastic mean free path. The size of the dots varied by orders of magnitude between the different experiments. 
On the theoretical side, addition spectra have attracted considerable interest. Restricting ourselves to the case of a large number, $N$, of electrons, interactions beyond the constant charging energy model give a Gaussian distribution of $\Delta_{2}(N)$ and increase the standard deviation of the addition spectra with respect to the constant charging energy approximation \cite{sivan,walker, berkovitz,ullmo, baranger,blanter}. These interactions are characterized by the ratio $r_s$ between the direct Coulomb and the kinetic Fermi energies.
The resulting standard deviation of the addition spectra includes a term due to the fluctuations of the single particle spectra and a term arising from the fluctuations of the charging energy $E_C=e^{2}/C_{\sum}$.  The former contribution is estimated for chaotic or diffusive dots from the random matrix theory (RMT): $\sigma_{2} \simeq 0.52\Delta_{1}$ \cite{sivan}. For small ($r_{s} \le 1$) or moderate values of $r_s$ \cite{walker}, the latter is typically estimated as $\sigma_{2}\simeq \Delta_1 {r_{s}\over \sqrt{g_{T}}}$, where $g_T \propto {e^{2}\over h} \times (k_{F} \ell) $ is the local dimensionless 2D conductance in the dot \cite{alhassid,blanter}. As a result $\sigma_{2}\simeq \Delta_{1} $, even after introducing spin \cite{baranger} or exchange \cite{alhassid} effects in the calculations. Smaller values of $g_T$ result in a larger $\sigma_{2}$ \cite{bonci}.
In the case of very large $r_s$ (negligible kinetic energy), one expects a Maxwell distribution, where the standard deviation scales as the charging energy ($\sigma_2 \propto E_C$) \cite{koulakov,koulakov2}. 

We have analyzed the distribution of the addition energies as a
function of the area of the quantum dot, restricting ourselves to the regime of large carrier densities in the dot. In contrast to previous experiments, we have studied the standard deviation of the addition spectra in low mobility silicon quantum dots obtained by resistive confinement from thin silicon-on-insulator (SOI) films \cite{jehl1}. We find that the fluctuations of the addition energy scale as the inverse area of the dot. The fluctuations are comparable to the mean level spacing for the one-particle spectrum, in agreement with most recent theories {\cite{baranger}.

The devices were made from boron doped ($10^{15}$ cm$^{-3}$) SOI (100) wafers. In order to ensure a low resistivity for the electrical contact to the silicon, the silicon film lying under the contacts is thick (70nm) and heavily doped. During doping, the active areas are protected from amorphization by a local thermal oxide layer. After its removal, we proceed to ion implantation (As ions above $10^{19}$ at.cm$^{-3}$) of the thin active areas (12 to 22 nm thick). 
A hybrid deepUV/Ebeam lithography combined with resist trimming is used to pattern silicon nanowires with width W in the range 20 to 400 nm and length $L_{f}$. After the silicon wires have been etched, a 4~nm-thick gate oxide was thermally grown before chemical vapor deposition of the in-situ doped poly-Si gate. Ebeam lithography was then used to pattern gates with length L down to 40 nm. The back end sequence follows a standard CMOS process. Figure \ref{fig:Figure1} shows a scanning electron micrograph (SEM) of the cross section of a typical wire. \\
%The samples are shaped from a SOI nanowire or constriction covered by a gate in the same way as in references \cite{chou,ishikuro,rokhinson}. 
The confinement of the dot is obtained by the part of highly resistive SOI wire away from the gate. There is no deliberate oxidation of the silicon film to create tunnel barriers as in the PADOX process \cite{padox}. We used neither fluctuations of the channel width to shape the quantum dot \cite{chou,ishikuro,rokhinson}, nor control gates to define the dot electrostatically. The latter procedure induces a deformation of the dot, which influences the addition spectra \cite{luscher,hackenbroich}. 
The area and shape of our dots is simply given by the overlap between the gate and the SOI film; the area can be as low as 3300 nm$^2$ and as high as 14400~nm$^2$. This results in values for $\Delta_1$, which are larger than in previous studies (see Table  \ref{table:Table1}) \cite{sivan,luscher,simmel,simmel2}. 

\begin{table}
     \caption{\label{table:Table1}Samples used in this study. $W$: wire width; 
     $L$: gate length; $L_{f}$: wire length; $d$: wire thickness; $\langle\delta V_{g}\rangle$: average peak spacing in gate voltage at $T$=4.2K; $\Delta_1$: one-particle mean level spacing. The dimensions are in nm.} 
     \begin{ruledtabular}
	 \begin{tabular}{c|c|c|c|c|c|cr}
	      & $W$ & $L$ & $L_{f}$ & $d$ & $\langle\delta V_{g}\rangle$ 
	     (mV) & $\Delta_1$ (meV) \\
	     \hline
         S1 & 70 & 40 & 200 & 12 & 3.8$\pm$0.3& 0.17 \\ %1CD7 Pl3
         S2 & 50 & 40 & 200 & 17 & 5.9$\pm$0.4& 0.19  \\ %1BD5 Pl12
	 S3 & 70 & 80 & 200 & 22 & 2.8$\pm$0.1& 0.07  \\ %1CD10 Pl21
	 S4 & 50 & 80 & 200 & 22 & 3.7$\pm$0.3& 0.09  \\ %OVL3D8 Pl21
	 S5 & 40 & 60 & 100 & 22 & 5.6$\pm$0.3& 0.13  \\ %OVL2D3 Pl21
	 S6 & 50 & 40 & 200 & 22 & 5.9$\pm$0.3& 0.17 \\ %1BD5 Pl21	 
	 S7 & 100 & 100 & 200 & 22 & 2.1$\pm$0.5& 0.04  \\ %OVL2AS10 Pl21
	 S8 & 70 & 40 & 200 & 22 & 6.7$\pm$0.2& 0.13 \\  %P21Pu42 S1C7 octobre2004
\end{tabular}
\end{ruledtabular}
\end{table}	 
A 2D electron accumulation layer is formed near the surface of the nanowire (see Figure \ref{fig:Figure1}).  The mobility was estimated from the sheet resistance of SOI films without gates. Under the gate it could be either enhanced, because of the protection provided by the gate or decreased by electrostatic disorder at the gate oxide/poly-Si interface. From $r_\square \simeq$ 4000 $\Omega$ (at room temperature as well as at $T$=4.2 K, for d=22nm) 
and the doping level, we estimate the mobility to $\mu \simeq 150$ cm$^{-2}$V$^{-1}$s$^{-1}$ and  $(k_{F} \ell) \simeq$ 3 -- 5.  $r_{s}$=$(a_\mathrm{Bohr}\sqrt{\pi n_{S}})^{-1}$ is 0.6 for $n_{S}$=$10^{13}$ cm$^{-2}$ (1.7 for $n_{S}$=$10^{12}$ cm$^{-2}$).\\ 
\begin{figure}
\begin{center}
\includegraphics[width=\linewidth]{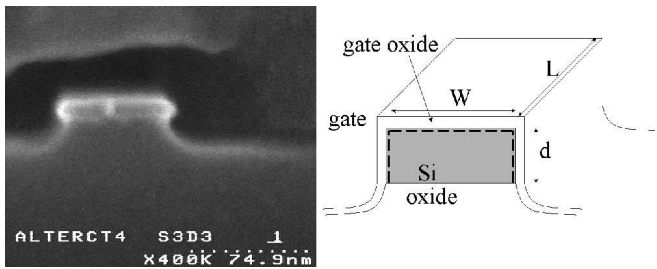}
\caption{Left: SEM picture of a specific morphological characterization structure: cross-section of a 12 nm thick and 60 nm wide silicon nanowire. Right: Layout of a sample showing the 2D electron gas (dash lines) of surface area $A=W\!\times\!L+2\!\times\!d\!\times\!L$. }
\label{fig:Figure1}
\end{center}
\end{figure}
Hundreds of samples with various geometries, doping and gate oxide thicknesses were integrated on 200 mm wafers. We  chose 20 samples to study at low temperature, which were cleaved and mounted in a He3/He4 dilution refrigerator with a working temperature range of 70 mK $< T <$ 4.2 K. An ac drain-source voltage $V_{DS}$ was applied and the $I_{DS}$ current measured with a standard lock-in technique. Lossy microcoaxes were used for RF filtering.
Due to the device variations (see Table~\ref{table:Table1}), we were able to adjust the access resistance to favor the 
observation of single-electron transport. Under these conditions the resistances varied little between $T$=300 K and $T$=4.2 K with values of the order of 100 k$\Omega$ measured at large gate voltage. This value is typical for having resistive confinement as demonstrated in \cite{zorin} and analyzed in \cite{nazarov}. The recorded features are remarkably stable, even after several cooling cycles.
The statistics are obtained by single $V_g$ scans covering several hundreds of peaks.

\begin{figure}
\begin{center}
\includegraphics*[width=\linewidth]{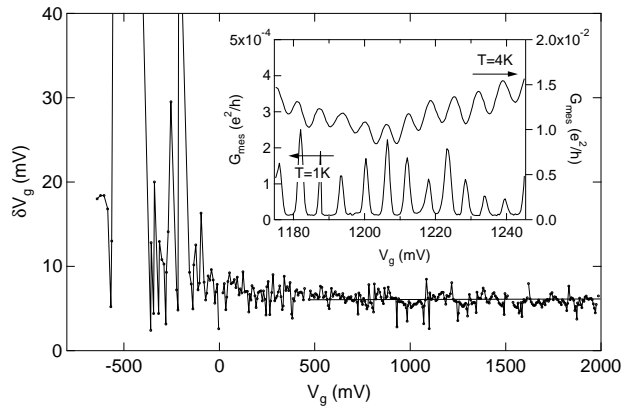}
\caption{Peak spacing $\delta V_{g}$ versus gate voltage $V_{g}$ of a Si-wire transistor (sample S6) at $T$=4.2 K. The inset shows typical conductance resonances as function of $V_{g}$ at $T$=1 K and $T$=4.2 K.
For the statistics only the high $V_{g}$ regime (many electrons) was considered, where $\delta V_{g}$ is approximately constant (slope of linear fit (line): $-10^{-4}$).}
\label{fig:Figure2}
\end{center}
\end{figure}

The inset of Figure~\ref{fig:Figure2} shows a detail of a typical drain-source conductance 
as a function of the gate voltage $V_{g}$. The spacings $\delta V_g$ between the peaks for the whole gate voltage range are shown below the inset.
Neglecting the first few dozen peaks, the mean value of 
the peak spacing in Figure \ref{fig:Figure2} is $\langle\delta V_{g}\rangle$=5.9$\pm$0.2 mV. An overview of $\langle\delta V_{g}\rangle$ for different wire geometries is given in Table \ref{table:Table1}. We have chosen samples with the same level of doping ($10^{19}$ As/cm$^3$) and the same gate oxide thickness ($t_{ox}$=4~nm).
The slope of a linear fit in the range 0.5 $< V_{g} <$ 2 V is of the 
order of 10$^{-4}$, indicating that the gate capacitance $C_{g}$ 
is little affected by the increasing number of electrons in the dot \cite{notebaranger}. 
This is in contrast with the observation of Ref. \cite{rokhinson}, where a shift of the period with gate voltage was attributed to an electrostatic coupling between two dots. We compared the measured values of $\langle\delta V_{g}\rangle$ with the expected values $\delta V_\mathrm{cal}=\frac{e}{C_g}$ for a planar capacitance $C_{g}=\epsilon_\mathrm{r}\epsilon_{0} A/t_{ox}$, where $A$ is the surface area of the 2D gas (see Figure \ref{fig:Figure1}) and $\epsilon_\mathrm{r}$ is the relative dielectric constant of the oxide.  
Figure~\ref{fig:Figure3} shows this ratio between measured and 
calculated mean peak spacings for the samples listed in Table~\ref{table:Table1}.
\begin{figure}
\begin{center}
\includegraphics*[width=\linewidth]{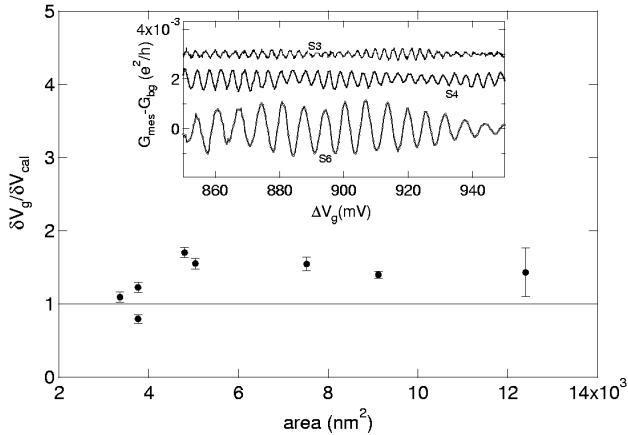}
\caption{Ratio of the measured ($\delta {V_{g}}$) and calculated ($\delta V_{cal}$) peak-to-peak distance for a flat capacitance $C=\epsilon_{0}\epsilon_\mathrm{r} A/t_{ox}$ ($\epsilon_\mathrm{r}$=3.9, area $A$, oxide thickness $t_{ox}$=4~nm). The inset shows measured oscillations after subtraction of the background in units of $e^{2}/h$ as a function of the gate voltage $V_g$ for three different areas (lowest curve: sample S6; middle curve: sample S4; highest curve: sample S3). Curves for S4 and S3 were shifted for clarity.}
\label{fig:Figure3}
\end{center}
\end{figure}
The ratios are close to 1, which shows that the dot size is, indeed, close to the area $A$ covered by the gate.
We estimate the mean level spacing $\Delta_{1}$=${\frac{\pi \hbar ^{2}}{2m^{*}A}}$ ($m^{*}$=$0.19m_e$ is the effective mass of the electron), taking into account the spin and band degeneracy for electrons on the (100) silicon surface \cite{fowler}. We compared this value with direct measurements of the excitation spectra in a sample of comparable size ($A=3300$ nm$^{2}$) but different doping from samples S1 to S7. For low carrier densities we find additional levels in the typical diamond shaped $V_{DS}-V_g$ dependence shown in the inset of Figure \ref{fig:Figure4}. The measurement of the level distances is in good quantitative agreement with the estimation of $\Delta_1$ considering that, due to the so-called Lifschitz tail, $\Delta_1$ is expected to be larger at lower densities, by roughly a factor of two or three \cite{fowler}.

For the analysis of the distribution of $\Delta_{2}$ we assumed a constant value for $\alpha$, as in previous works, although it decreases slightly with $V_g$.
This variation is enhanced at low $V_g$ (in the first peaks region omitted from the statistics) because of the reduction of the source and drain capacitances.  
As usual in MOS-Field Effect Transistors near threshold, the capacitances between the channel and source/drain and body, respectively, increase with the gate voltage, while the gate capacitance remains constant \cite{sze}.
$\alpha$ is obtained from the slopes of the Coulomb diamonds at intermediate gate voltages and by fitting the temperature dependence of the resonances at high gate voltage. 
$\alpha \simeq 0.4$ is found for all samples from the analysis of the Coulomb diamonds.
 We also note that $\alpha$ can fluctuate from resonance to resonance around its mean value, an effect neglected here which deserves a more systematic study.
\begin{figure}
\begin{center}
\includegraphics*[width=\linewidth]{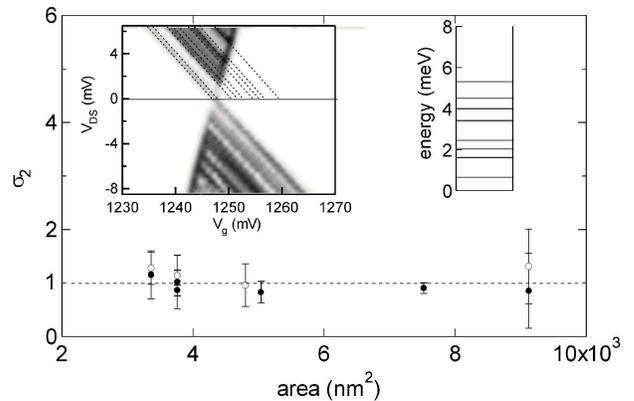}
\caption{Fluctuations of the peak spacing distributions $\sigma_2$, normalized to the mean one-particle energy level spacing $\Delta_1$, versus area $A$ (filled dots: 4.2~K, open dots: 1~K). The inset shows the typical $V_{DS}-V_{g}$ diamonds of the first conductance peak in a $A$=3300~nm$^{2}$ sample. The one-particle levels appear as additional lines in the diamonds. Projecting along the $V_{g}$ axis and taking $\alpha=0.4$ gives the measured excitation spectrum, with a mean value $\Delta_1=0.6\pm0.2$ meV, in good agreement with the calculated values of Table \ref{table:Table1} (see text).}
\label{fig:Figure4}
\end{center}
\end{figure}
Figure \ref{fig:Figure5} shows the distribution of the peak spacings in gate voltage $\delta V_g$ for three different areas. As shown in the inset of Figure \ref{fig:Figure3}, the distributions shift to higher mean values of $\delta V_g$ as the sample size decreases. At the same time a broadening of the distribution can be observed. The measurements were usually done at $T$=4.2 K, i.e. $k_{B}T\gtrsim\Delta_{1}$. A comparison with measurements at lower temperature, $T$=1~K and $T$=300~mK (effective minimal electronic temperature), has shown that the addition spectra is sligthly modified: $\sigma_2$ increases at lower temperature, but this increase lies within the error bars of the $T$=4.2 K statistics (see inset of Figure \ref{fig:Figure5}).
As in previous reports we observe a Gaussian distribution for the addition energies with a variance $\sigma_2$, which is comparable to $\Delta_1$. The Gaussian shape contradicts the
Constant Interaction approximation within the RMT model, where the ground state energy is the sum of the charging energy and the energies of the occupied single particle states \cite{alhassid}. To compare our results with theory \cite{baranger}, we plotted $\sigma_2$ normalized to $\Delta_1$ 
as a function of the dots' area (see Figure \ref{fig:Figure4}). Note that the simulation in Ref.\cite{baranger} uses parameters comparable to ours: $r_{S} \simeq 1.5$, $g_{T} \simeq 4$ and $N \simeq 200$. Due to the smaller size of our dots, $\Delta_1$ is order of magnitudes larger than in \cite{patel,simmel}. Alternatively, the inset of Figure \ref{fig:Figure5} shows the standard deviation $\sigma$ of the frequently used normalization $(\delta V_g-\langle\delta V_g\rangle)/\langle\delta V_g\rangle$, obtained in Gaussian fits. The standard deviation increases when scaling down the area of the dots. $\sigma$ reaches about 10 \% for the smallest sample ($A$=3300 nm$^2$). This value is close to the ratio $\Delta_1\sim$0.19 meV over the value of the addition energy, $\Delta_2=e\alpha\langle\delta V_g \rangle$=2.4 meV: $\Delta_1/\Delta_2$=0.08. The normalization of $\sigma_2$ with $\Delta_1$ (Figure \ref{fig:Figure4}) leads to a constant $\frac{\sigma_2}{\Delta_1}$ (here $\simeq$1), which proves that $\sigma_2$ scales with $\Delta_1$ in our samples.
\begin{figure}
\begin{center}
\includegraphics*[width=\linewidth]{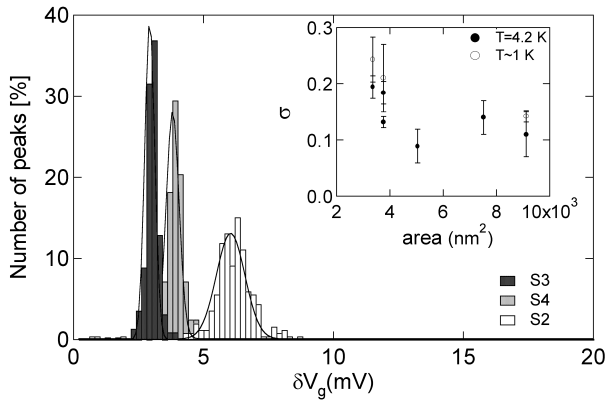}
\caption{Relative peak spacing distribution $P(\delta V_{g})$ for three different dots' areas at 4.2~K. The solid lines are the gaussian fits used to extract the standard deviation. The shift towards larger values of $\delta V_{g}$ for decreasing areas is clearly visible. The distribution becomes larger when the sample area decreases. This enhancement of the fluctuations in smaller samples is shown in the inset, where the standard deviation $\sigma$ of the normalized fluctuation $(\delta V_g-\langle\delta V_g\rangle)/\langle\delta V_g\rangle$ is plotted against the surface area of the sample.}
\label{fig:Figure5}
\end{center}
\end{figure}
At low gate voltages it is obvious from Figure \ref{fig:Figure2} that the fluctuations in $\sigma_2$ are greatly enhanced. We assumed that due to the low mobility of our samples and the low carrier concentration at low gate voltages, the electrons cross a mobility edge and the dot enters the localized regime. As the added electrons and their screening clouds are both localized on different sites, distant from almost zero to the diameter of the dot, the relative fluctuations of the addition spectra can reach 100\% as explained in Ref.\cite{koulakov}. 

We presented a systematic study of the size dependence of the addition spectra in low mobility and ultra small silicon dots. Our dots are very similar to ultimate silicon MOSFETs, different from those used in previous systematic experiments. The small size implicates a one-particle mean level spacing of the order of a few Kelvin, which is much larger than in previous experiments. Due to the chosen fabrication technique, the dot shape is not modified by the gate voltage, which only modifies the concentration of electrons in the dot. We have found an excellent agreement with theories in the same range of parameters, both for the expected values and the size dependency of the fluctuations. The presented analysis is restricted to the high concentration range of carriers. Experiments investigating the low density case, where the dots undergo the metal-insulator transition, are in progress. 

\begin{acknowledgments}
This work was partly supported by the European Commission under the frame of the Network of Excellence "SINANO" (Silicon-based Nanodevices, IST-506844).
\end{acknowledgments}


\begin{thebibliography}{29}
\bibitem{notebaranger} Nevertheless we take into account this variation of $C_g$ to obtain the standard deviation $\sigma_2$.
\end{thebibliography}
\end{document}